\documentclass[preprint]{raa}            % referee version: for submission
\usepackage{graphicx,times}             %for PS/EPS graphics inclusion, new
\usepackage{natbib}
\usepackage{amssymb,amsmath}
\bibpunct{(}{)}{;}{a}{}{,}

\usepackage[final=true,dvipdfm=true,pagebackref=true]{hyperref}
\hypersetup{colorlinks = true, linkcolor = green, anchorcolor = red, citecolor = blue, filecolor = red, pagecolor = red, urlcolor = red}

\begin{document}

   \title{One large glitch in PSR B1737$-$30 detected with the TMRT
%\footnotetext{$*$ Supported by the National Natural Science Foundation of China.}
}
%   \subtitle{I. Place Your Subtitle Here}

   \volnopage{Vol.0 (20xx) No.0, 000--000}      %%preserved for Editor. DOn't remove!
   \setcounter{page}{1}          %%starting page, preserved for Editor. DOn't remove!

   \author{Jie Liu
      \inst{1,4}
   \and Zhen Yan
      \inst{1,6}
   \and Jian-Ping Yuan
      \inst{2}
   \and Ru-Shuang Zhao
      \inst{1,5}
   \and Zhi-Peng Huang
      \inst{1,4}
   \and Xin-Ji Wu
      \inst{3}
   \and Na Wang
      \inst{2}
   \and Zhi-Qiang Shen
      \inst{1,4,6}
   }

   \institute{Shanghai Astronomical Observatory, Chinese Academy of Sciences, Shanghai 200030, China; {\it yanzhen@shao.ac.cn}\\
        \and
             Xinjiang Astronomical Observatory, Chinese Academy of Sciences, Xinjiang 830011, China\\
        \and
             School of Physics, Peking University, Beijing 100871, China \\
        \and
             School of Physical Science and Technology, ShanghaiTech University, Shanghai 201210, China\\
        \and
             University of Chinese Academy of Sciences, Beijing 100049, China \\
        \and
             Key Laboratory of Radio Astronomy, Chinese Academy of Sciences, Nanjing 210008, China\\
\vs\no
   {\small Received~~20xx month day; accepted~~20xx~~month day}}

\abstract{ One large glitch was detected in PSR B1737$-$30 using data spanning from MJD 57999 to 58406 obtained with the newly built Shanghai Tian Ma Radio Telescope (TMRT). The glitch took place at the time around MJD 58232.4 when the pulsar underwent an increase in the rotation frequency of $\Delta \nu$ about 1.38$\times 10^{-6}$ Hz, corresponding to a fractional step change of $\Delta \nu / \nu$ $\thicksim$ 8.39$\times 10^{-7}$. Post$\textrm{-}$glitch $\nu$ gradually decreased to the pre$\textrm{-}$glitch value. The frequency derivative was observed to undergo a step change of about $-$9$\times 10^{-16}$ s$^{-2}$. Since July 1987, there are 36 glitches already reported in PSR B1737$-$30 including this one. According to our analysis, the glitch size distribution is well described by the power law with index of 1.13. The distribution of the interval between two adjacent glitches (waiting time $\Delta T$) follows a Poissonian probability density function. For PSR B1737$-$30, the interval is prone to be long after a large glitch. But no correlation is found between glitch size and the interval since previous glitch.
\keywords{stars: neutron --- pulsars: general --- pulsars: individual: B1737$-$30}
}

   \authorrunning{Jie Liu et al. }            %author_head in even pages
   \titlerunning{One Glitch in PSR B1737$-$30}  % title_head in odd pages

   \maketitle
%% The author head (on even pages) and the title head (on odd pages) will be
%% automatically extracted from \author{} and \title{}. Whenever the title is too long,
%% you will be asked to supply a shorter one by inserting either \authorrunning{} or
%% \titlerunning{} before \maketitle. Anyway, you can specify your own heads.
%%
%%
%% Note: In the following text body of your manuscript, please note several differences from
%%       other major journals:
%% (1) \subsection{Please Capitalize the First Letter of Each Notional Word in Subsection Title}
%% (2) Please Capitalize the First Letter of Each Notional Word in all tables' captions

%
%________________________________________________ sections below
%
\section{Introduction}
\label{sect:intro}

In general, pulsars rotate with high stability, making it possible to predict arrival time of each pulse over long time. However, two kinds of timing irregularities have been detected in pulsar rotation evolutions: the timing noise and the glitch. The timing noise is a kind of long$\textrm{-}$term stochastic fluctuation in residuals. It is related to pulsar characteristic age $\tau_{\rm c}$ = $P/(2\dot{P}$) and the spin$\textrm{-}$down rate $\left|\dot{\nu} \right| $ \citep{hlk10}. By comparison, the glitch is a sudden change in the rotation frequency.

Since the first glitch was detected in Vela pulsar (B0833$-$45) \citep{rm69,rd69}, there are about 520 glitches detected in 180 pulsars \citep{man18}. Almost all frequency jumps ($\Delta \nu$) caused by glitches are positive except two negative cases in PSRs J1522$-$5735 \citep{pga+13} and J2301$+$5852 \citep{akn+13}. The distribution of $\Delta \nu$ widely ranges from $10^{-11}$ to $10^{-4}$ Hz \citep{elsk11,fer17}. The smallest glitch was detected in PSR J0631$-$0200 with a $\Delta \nu / \nu$ about 2.5$\times 10^{-12}$ \citep{mjs+16}, and the largest glitch was observed in PSR J1718$-$3718 with a $\Delta \nu / \nu$ about 3.325$\times 10^{-5}$ \citep{mh11}. Post$\textrm{-}$glitch rotation frequency relaxes back towards the pre$\textrm{-}$glitch value in most cases. Exponential processes are observed in the relaxation process for some glitches. The time scale of relaxation evidently differs from one glitch to another ranging from minutes to years \citep{lkb+96,dlm07}.

Half century has passed since the first pulsar glitch was detected, but glitch events are still not well understood. The vortex model \citep{ai75,apas84,hm15} is commonly used to explain the internal mechanism of glitch. In this scenario, neutrons in the pulsar interiors are assumed to be superfluid \citep{bpp69}. Vortices are pinned to nuclei in solid crust or the core of pulsars and limited to move outward due to the interaction with ions in the neutron star, so the superfluid cannot loose vorticity to spin down and rotates faster than crust. Once Magnus force frees vortices, the angular momentum is transferred rapidly from superfluid to crust, giving the rise of crust rotation. Soon after glitch, the vortices are repinned to other regions, causing the relaxation of rotation frequency towards initial value. There is another kind of timing irregularity named slow glitch, where $\nu$ gradually increases after glitch and the $\left|\dot{\nu} \right| $ undergoes a quick decrease accompanied by an exponential recovery \citep{zww+04,sha05}. It is predicted that the temperature fluctuation of neutron star will cause the gradual increase in rotation frequency \citep{gre79a}. Slow glitch happens if the local temperature in inner crust increases suddenly \citep{le96}. The decrease in $\left|\dot{\nu} \right| $ may be a response to the decrease in the braking torque \citep{sha05}.

\begin{table}[ht]
\footnotesize{
\caption{Parameters of PSR B1737$-$30.}
\label{Tab:table1}
\centering
\begin{tabular}{c c c c c c c c c c }
\hline
\hline
\multicolumn{2}{c}{Name} &
\multicolumn{1}{c}{RA} &
\multicolumn{1}{c}{DEC} &
\multicolumn{1}{c}{$P$} &
\multicolumn{1}{c}{$\dot{P}$} &
\multicolumn{1}{c}{$\tau_c$} &
\multicolumn{1}{c}{DM}  &
\multicolumn{1}{c}{$S_{1400}$}
\\
\multicolumn{1}{c}{B1950} &
\multicolumn{1}{c}{J2000} &
\multicolumn{1}{c}{(h \ m \ s)}  &
\multicolumn{1}{c}{(d \ m \ s)} &
\multicolumn{1}{c}{(s)} &
\multicolumn{1}{c}{($10^{-13}$ s/s)}  &
\multicolumn{1}{c}{(kyr)} &
\multicolumn{1}{c}{(pc cm$^{-3}$)} &
\multicolumn{1}{c}{(mJy)}
\\
\hline
B1737$-$30 & J1740$-$3015 & 17:40:33.82 & $-$30:15:43.5 & 0.60688662425 & 4.66124 & 20.6 & 151.96 & 8.9
\\
\hline
\end{tabular}
\\
Note: All these parameters are referenced from ATNF pulsar database ({\it http://www.atnf.csiro.au/research/pulsar/psrcat/}). $S_{1400}$ is the flux density at 1.4 GHz.
}
\end{table}

PSR B1737$-$30 was detected in the high$\textrm{-}$radio$\textrm{-}$frequency survey at Jodrell Bank \citep{cl86} in 1986. Its rotation period is 0.607 s and the period derivative is about 4.66$\times 10^{-13}$ s/s \citep{ywml10}, suggesting a young characteristic age $\tau_{\rm c}$ of 20.6 kyr. The parameters of PSR B1737$-$30 are listed in Table~\ref{Tab:table1}. PSR B1737$-$30 exhibits frequent glitch events with 35 glitches reported during MJD 46991 (July 15, 1987) and 57499 (April 21, 2016). The timing properties of this pulsar were also monitored by the Shanghai Tian Ma Radio Telescope (TMRT) which is a newly built radio telescope with the diameter of 65 m. In this paper, we present one large glitch detected by the TMRT. The structure of this paper is organized as below. Observations together with data analysis are described in section~\ref{sect:Observation analysis}. Detail results are shown in section~\ref{sect:results}. The discussion and a short conclusion are presented in section~\ref{sect:discussion} and~\ref{sect:conclusion}.

%% Authors can give a citation as 'Michel et al. 1992'.
%% You may also use \cite, \citep and \citet for citation, and use Table~1 or Figure~1
%% and so forth. Using \ref and \label for cross-references of Tables/Figures
%% is a good way in adjusting/adding/removing text, tables or figures.

\section{Observations and Analysis}
\label{sect:Observation analysis}

Timing observations of PSR B1737$-$30 were carried out at the wavelength of 13 cm with the TMRT between MJD 57999 (September 3, 2017) and 58406 (October 15, 2018), using the S$\textrm{-}$band cryogenically cooled, dual$\textrm{-}$polarization receiver. The effective frequency coverage of the receiver ranges from 2.2 to 2.3 GHz \citep{ysm+18}. The full bandwidth is divided into channels with the typical width of 1 MHz for the convenience to remove the dispersion effect and the radio frequency interference (RFI). Data sampling and recording are accomplished using the digital backend system (DIBAS) with time resolution of 40.96 $\mu$s \citep{ysw+17}. The incoherent dedispersion on$\textrm{-}$line folding observation mode was used in the timing observations with subintegration time of 30 s \citep{ysw+15}. The folding parameters are obtained from ATNF pulsar catalogue \citep{mhth05} \footnote{http://www.atnf.csiro.au/research/pulsar/psrcat/}. The observation data are written out as 8$\textrm{-}$bit {\sc PSRFITS} files. Each period is divided into 1024 phase bins. Duration of observations were mostly from 10 to 20 min, depending on observation conditions (e.g.,weather, RFI, etc.).

\begin{figure}[h]
\begin{center}
\begin{tabular}{cc}
\resizebox{0.64\hsize}{!}{\includegraphics[angle=0]{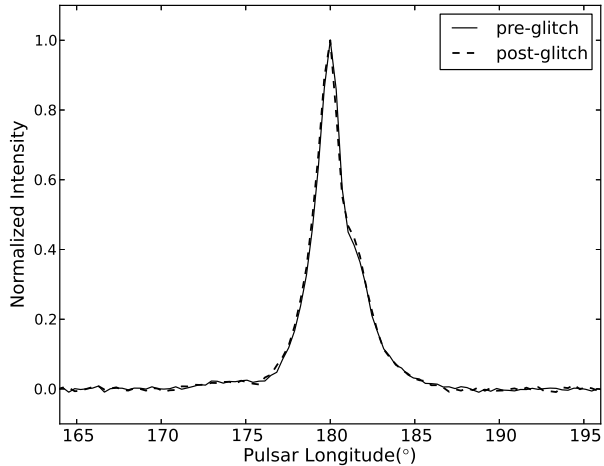}}\\
\end{tabular}
\end{center}
\caption{The integrated normalized pulse profiles of pre$\textrm{-}$ and post$\textrm{-}$glitch.}
\label{fg:msRAA-2018-0260R1fig1}
\end{figure}

\begin{figure}[h]
\begin{center}
\begin{tabular}{cc}
\resizebox{0.64\hsize}{!}{\includegraphics[angle=0]{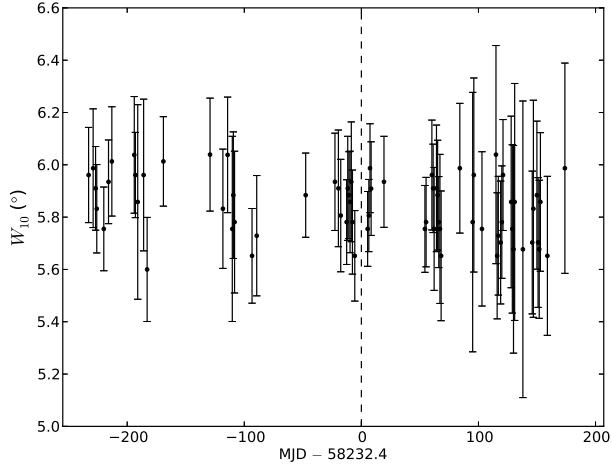}}\\
\end{tabular}
\end{center}
\caption{Distribution of $W_{10}$ in unit of degree. The dashed vertical line implies the epoch of glitch.}
\label{fg:msRAA-2018-0260R1fig2}
\end{figure}

In the pulsar timing observations, the time was kept with local hydrogen atomic clock corrected to GPS. Data reduction and analysis were performed using the {\sc PSRCHIVE} \citep{hvm04} and the {\sc TEMPO2} \citep{hob12}. Data from all channels and subintegrations were scrunched together to get the mean pulse profile for each single observation. Pre$\textrm{-}$ and post$\textrm{-}$glitch pulse profiles were integrated separately. The integrated normalized pulse profiles are shown in Fig~\ref{fg:msRAA-2018-0260R1fig1}. There is no obvious difference between widths of pulse profiles before and after glitch (Fig~\ref{fg:msRAA-2018-0260R1fig2}). Local pulse times of arrival (TOAs) were generated through the cross$\textrm{-}$correlation of observed pulse profiles with a high signal to noise ratio (SNR) pulse profile (template). They were converted to TOAs at the Solar$\textrm{-}$system barycenter with the Jet Propulsion Laboratories DE405 ephemeris \citep{sta98b}. TOA errors are mostly in the range of 20$\textrm{-}$50 $\mu$s. The pulse phase $\phi$ at Solar$\textrm{-}$system barycenter given by model is a Taylor series which can be described as a function of time $t$ as below:
\begin{equation}
  \phi(t) = \phi_{\rm 0} + \nu(t-t_{\rm 0}) + \frac{1}{2}\dot{\nu}(t-t_{\rm 0})^{2} + \frac{1}{6}\ddot{\nu}(t-t_{\rm 0})^{3} + \cdots ,
\label{eq:equation1}
\end{equation}
where $\phi_{\rm 0}$, $\nu$, $\dot{\nu}$, $\ddot{\nu}$ are the phase at $t_{\rm 0}$, rotation frequency with its first and second time$\textrm{-}$derivatives, respectively.

Post$\textrm{-}$glitch frequency typically relaxes back towards pre$\textrm{-}$glitch value in the form of:

\begin{equation}
  \nu(t) = \nu_{\rm 0}(t) + \Delta \nu_{\rm p} + \Delta \dot{\nu}_{\rm p}t + \Delta \nu_{\rm d}e^{-t/\tau_{\rm d}},
\label{eq:equation2}
\end{equation}

\begin{equation}
  \dot{\nu}(t) = \dot{\nu}_{\rm 0}(t) + \Delta \dot{\nu}_{\rm p} +  \Delta \dot{\nu}_{\rm d}e^{-t/\tau_{\rm d}},
\label{eq:equation3}
\end{equation}
where $\Delta \nu_{\rm p}$, $\Delta \dot{\nu}_{\rm p}$, $\tau_{\rm d}$ and $\Delta \nu_{\rm d}$ are permanent changes in $\nu$ and $\dot{\nu}$ relative to pre$\textrm{-}$glitch values, time constant and amplitude of exponential decay, respectively. The total frequency increment caused by glitch is
\begin{equation}
  \Delta \nu =  \Delta \nu_{\rm p} + \Delta \nu_{\rm d},
\label{eq:equation4}
\end{equation}

The degree of recovery can be described by the parameter $Q$:
\begin{equation}
  Q =  \Delta \nu_{\rm d} / \Delta \nu ,
\label{eq:equation5}
\end{equation}

Results of PSR B1737$-$30 were obtained using observation data ranging from MJD 57999 to 58406. Timing residual is the difference between barycentric arrival time and the predicted TOA, which randomly distributes around zero if the rotation of pulsar is well described by the simple slow$\textrm{-}$down model. Once a glitch happens, the rotation suddenly speeds up (or down), causing the earlier (or later) arrival of pulses than predicted by model. So the timing residuals will obviously decrease towards negative (or positive) value. Timing residuals of PSR B1737$-$30 in the left panel of Fig~\ref{fg:msRAA-2018-0260R1fig3} show an obvious downward trend after MJD 58232.4 (April 24, 2018), suggesting a large glitch at that time. In order to confirm whether this glitch event is caused by improperly corrected jump and drift of clocks or not, we did further timing analysis of the millisecond pulsar B1937$+$21 \citep{bkh+82}, which was also monitored at TMRT with same setups. The timing residuals are shown in the right panel of Fig~\ref{fg:msRAA-2018-0260R1fig3}. As no obvious change was found in residuals of PSR B1937$+$21 around MJD 58232.4, the distinct change in the residuals of PSR B1737$-$30 was caused by the glitch. Since the data interval around glitch is about 11 days, the final glitch epoch was estimated in two steps. Firstly, it was estimated as the middle point of the interval. Then, we fitted all the glitch parameters using {\sc TEMPO2} while changing the glitch epoch. The final epoch value was selected when chi$\textrm{-}$square (${\chi}^2$) became minimum. The error of glitch epoch was taken as the region of epochs corresponding to $\Delta \chi^{2}$ $\leq$ 1 from the minimum. Pre$\textrm{-}$ and post$\textrm{-}$glitch frequency parameters were revealed by fitting $\nu$, $\dot{\nu}$, $\ddot\nu$ with data before and after glitch separately. In order to know the evolution of $\nu$ and $\dot{\nu}$ around glitch, we figured out frequency residuals at various epochs. They were obtained by fitting Equation~\ref{eq:equation1} (omitting the $\ddot\nu$ term) over a series of overlapping data sections (Table~\ref{Tab:table2}). The time scales of data sections range from 13 to 80 d. The epoch of each fit was set to be the middle date of the data section.

\begin{table}[ht]
\footnotesize{
\caption{Timing solutions of data sections.}
\label{Tab:table2}
\centering
\begin{tabular}{c c l c c }
\hline
\hline
\multicolumn{1}{c}{Data section} &
\multicolumn{1}{c}{Epoch} &
\multicolumn{1}{c}{$\nu$} &
\multicolumn{1}{c}{$\dot{\nu}$} &
\multicolumn{1}{c}{Numbers of}
\\
\multicolumn{1}{c}{(MJD)} &
\multicolumn{1}{c}{(MJD)} &
\multicolumn{1}{c}{(s$^{-1}$)}  &
\multicolumn{1}{c}{(10$^{-12}$ s$^{-2}$)}  &
\multicolumn{1}{c}{TOAs}
\\
\hline
 57999$\textrm{-}$58012 & 58006 & 1.6474050385(1)  & $-$1.2639(7) & 5 \\
 58012$\textrm{-}$58039 & 58025 & 1.6474050387(1)  & $-$1.2647(4) & 5 \\
 58038$\textrm{-}$58063 & 58050 & 1.64740503873(4) & $-$1.2638(1) & 6 \\
 58063$\textrm{-}$58124 & 58113 & 1.64740503875(5) & $-$1.2641(8) & 7 \\
 58114$\textrm{-}$58185 & 58132 & 1.64740503876(6) & $-$1.2638(6) & 8 \\
 58139$\textrm{-}$58220 & 58199 & 1.64740503883(4) & $-$1.2641(3) & 7 \\
 58184$\textrm{-}$58227 & 58225 & 1.6474050387(1)  & $-$1.2649(7) & 10 \\
 58237$\textrm{-}$58252 & 58244 & 1.6474064190(2)  & $-$1.2668(8) & 5 \\
 58286$\textrm{-}$58301 & 58293 & 1.6474064104(1)  & $-$1.2643(9) & 10 \\
 58316$\textrm{-}$58353 & 58335 & 1.6474064051(1)  & $-$1.2649(2) & 10 \\
 58347$\textrm{-}$58370 & 58350 & 1.6474064033(2)  & $-$1.2653(7) & 11 \\
 58360$\textrm{-}$58385 & 58373 & 1.6474064013(1)  & $-$1.2655(6) & 11 \\
 58363$\textrm{-}$58406 & 58395 & 1.6474063999(3)  & $-$1.2652(1) & 10 \\
\hline
\end{tabular}
\\
}
\end{table}

\section{Results}
\label{sect:results}

\begin{table}[ht]
\centering
\footnotesize{
\caption{Timing solutions and glitch parameters.}
\label{Tab:table3}
\begin{tabular}{l c c}
\hline
\hline
\multicolumn{1}{c}{Parameter} &
\multicolumn{1}{c}{Pre$\textrm{-}$glitch} &
\multicolumn{1}{c}{Post$\textrm{-}$glitch}
\\
\hline
$\nu$ (Hz)                         & 1.64740503872(5)       & 1.647406475(2) \\
$\dot{\nu}$ (10$^{-12}$ s$^{-2}$)  & $-$1.26397(2)          & $-$1.2678(2) \\
$\ddot{\nu}$ (10$^{-23}$ s$^{-3}$) & $-$1.6(2)              & 7.8(7) \\
Frequency epoch (MJD)              & 58113                  & 58322 \\
Data span (MJD)                    & 57999$\textrm{-}$58227 & 58237$\textrm{-}$58406 \\
TOA numbers                        & 32                     & 38 \\
\\
GlitchEpoch (MJD)                          & \multicolumn{2}{c}{58232.4(4)} \\
$\Delta \nu$ ($10^{-9}$ Hz)                & \multicolumn{2}{c}{1381.7(8)} \\
$\Delta \nu / \nu$ ($10^{-9}$)             & \multicolumn{2}{c}{838.7(5)} \\
$\Delta \dot{\nu}$ ($10^{-16}$ s$^{-2}$)   & \multicolumn{2}{c}{$-$9.0(4)} \\
$\Delta \dot{\nu} / \dot{\nu}$ ($10^{-3}$) & \multicolumn{2}{c}{0.71(3)} \\
$\Delta \nu_{\rm d}$ ($10^{-9}$ Hz)        & \multicolumn{2}{c}{9.5(6)}\\
$\tau_{\rm d}$ (d)                         & \multicolumn{2}{c}{71(6)}\\
RMS residual ($\mu$s)                      & \multicolumn{2}{c}{201.27} \\

\hline
\end{tabular}
\\
}
\end{table}

\begin{figure}[h]
\begin{center}
\begin{tabular}{cc}
\resizebox{0.5\hsize}{!}{\includegraphics[angle=0]{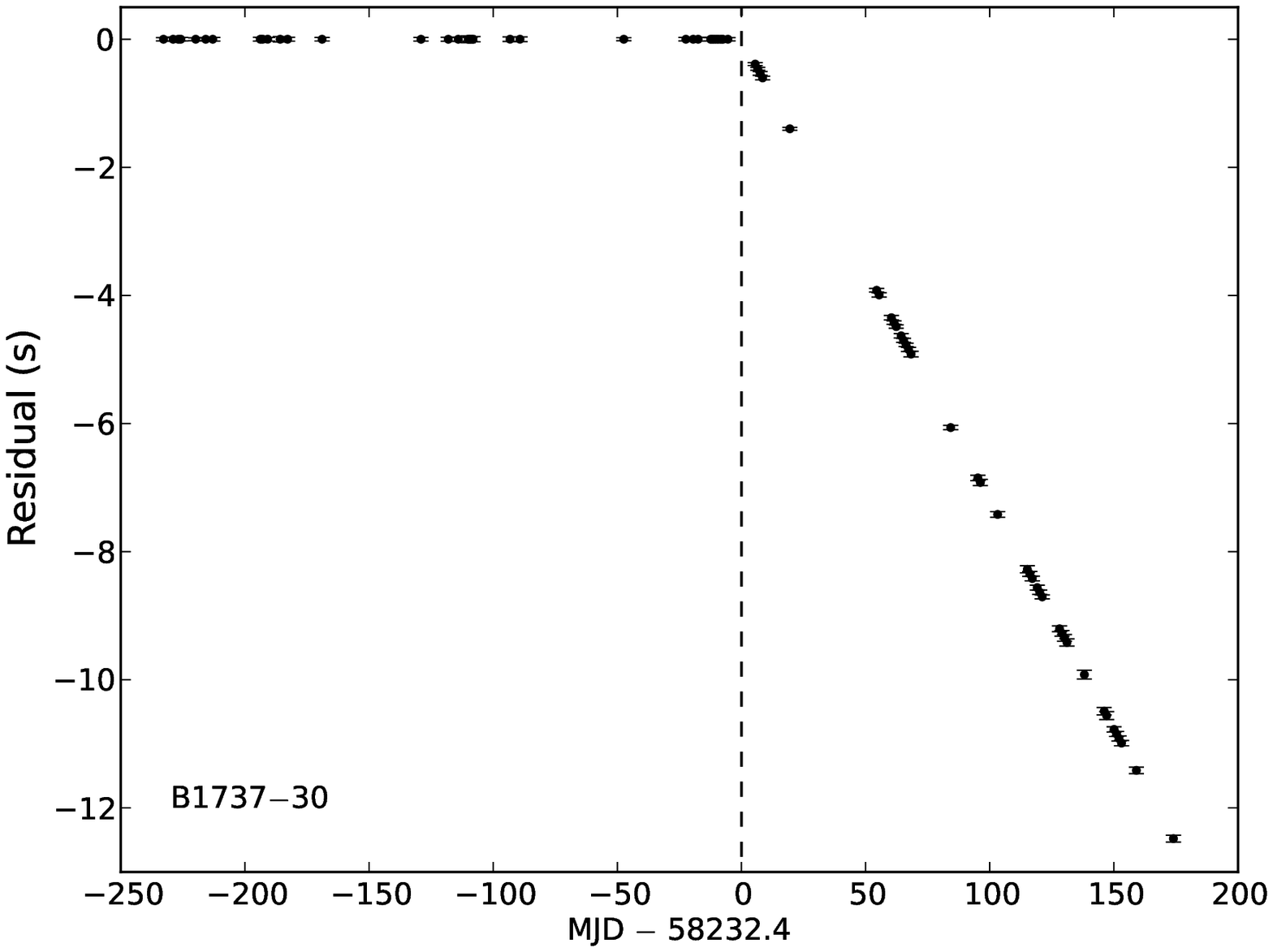}} &
\resizebox{0.5\hsize}{!}{\includegraphics[angle=0]{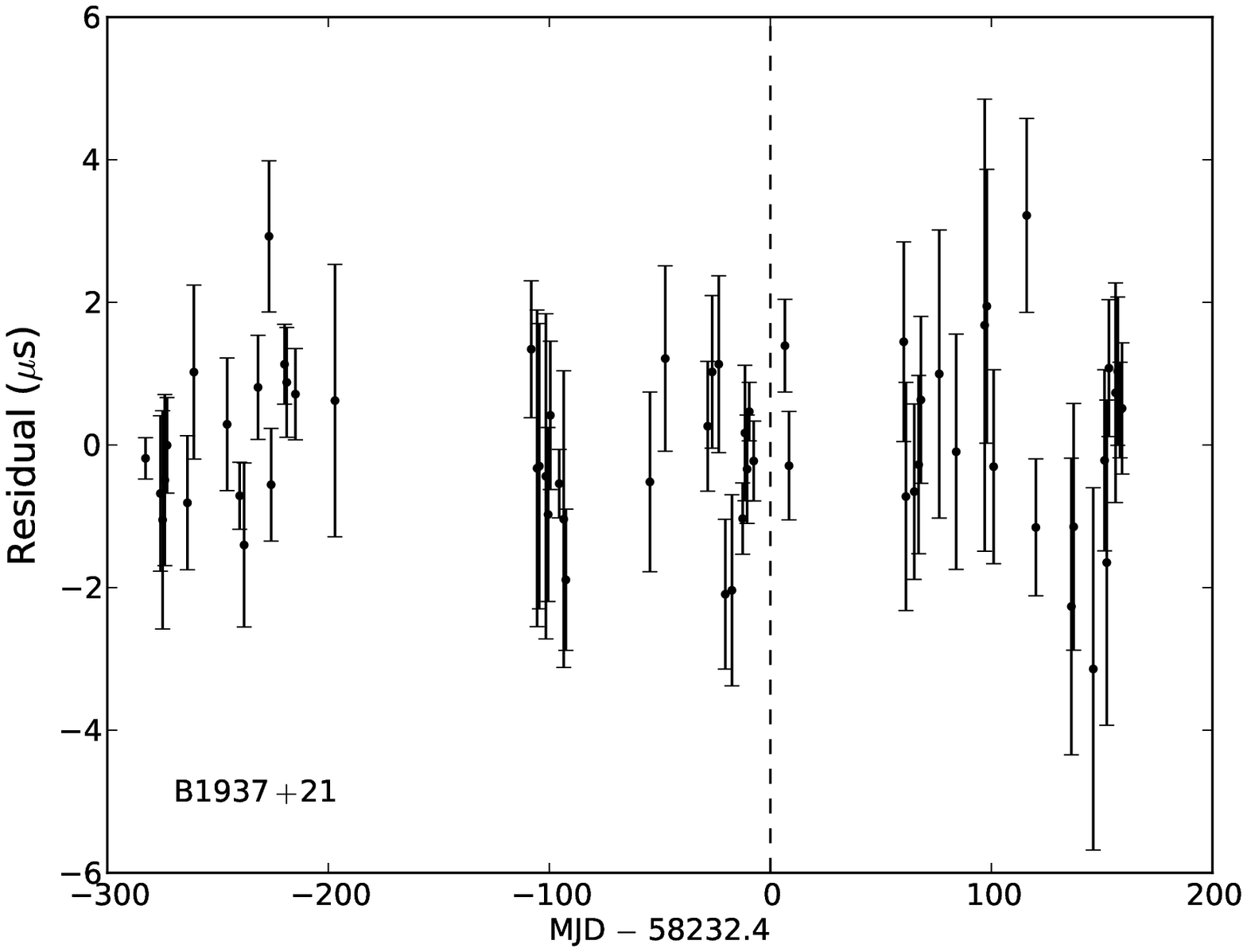}}\\
\end{tabular}
\end{center}
\caption{Timing residuals of PSRs B1737$-$30 (left) and B1937$+$21 (right). The dashed vertical line implies the epoch of the glitch.}
\label{fg:msRAA-2018-0260R1fig3}
\end{figure}

\begin{figure}[h]
\begin{center}
\begin{tabular}{cc}
\resizebox{0.65\hsize}{!}{\includegraphics[angle=-90]{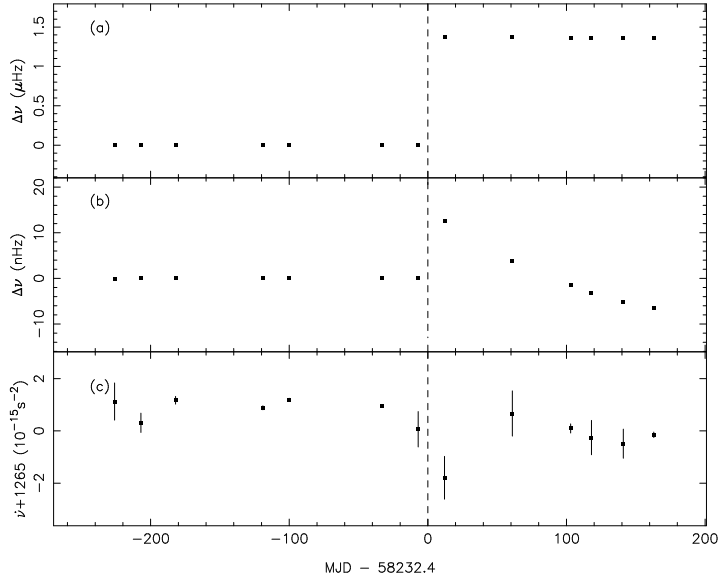}}&
\end{tabular}
\end{center}
\caption{Frequency variations of PSR B1737$-$30 relative to pre$\textrm{-}$glitch solutions. (a): Residuals of $\Delta \nu$ after subtracting pre$\textrm{-}$glitch spin$\textrm{-}$down model; (b): $\Delta \nu$ with mean values removed separately before and after glitch; (c): The evolution of frequency derivative $\dot{\nu}$ corresponds to an initial value of $\left|\dot{\nu} \right| $. The dashed vertical line implies the epoch of the glitch.}
\label{fg:msRAA-2018-0260R1fig4}
\end{figure}

The timing solutions of PSR B1737$-$30 around MJD 58232.4 are listed in Table~\ref{Tab:table3} together with glitch parameters. The glitch parameters were obtained by fitting Equation~\ref{eq:equation2} and~\ref{eq:equation3} with {\sc TEMPO2}. Timing residuals relative to the pre$\textrm{-}$glitch rotation model are shown in Fig~\ref{fg:msRAA-2018-0260R1fig3} (left panel). The residuals between MJD 57999 and 58227 randomly distributed around zero, implying that the model fits well. After the occurrence of glitch, residuals continuously decreased towards negative value. The root mean square (RMS) residual is 201.27 $\mu$s after subtracting the fitted glitch model, which corresponds to $\thicksim$ 0.001 turns. Evolution behaviours of $\nu$ and $\dot{\nu}$ are shown in Fig~\ref{fg:msRAA-2018-0260R1fig4}. The variations in $\nu$ with pre$\textrm{-}$glitch model subtracted are plotted in panel (a). It shows a remarkable increment about 1.38$\times 10^{-6}$ Hz, corresponding to a $\Delta \nu / \nu$ $\thicksim$ 8.39$\times 10^{-7}$. This increment is also demonstrated by pre$\textrm{-}$ and post$\textrm{-}$glitch $\nu$ values listed in Table~\ref{Tab:table3}. After subtraction of the mean values separately for pre$\textrm{-}$ and post$\textrm{-}$glitch $\nu$, it is obviously shown in panel (b) that $\nu$ exponentially decayed towards initial value after glitch. The time constant of the exponential decay $\tau_{\rm d}$ was fitted to be 71 d. The amplitude of $\Delta \nu_{\rm d}$ is about 9.5$\times 10^{-9}$ Hz, possibly implying a small value of recovery index $Q$. Panel (c) shows that the spin$\textrm{-}$down rate $\left|\dot{\nu} \right| $ underwent an increase about 9$\times 10^{-16}$ s$^{-2}$, corresponding to a $\Delta \dot{\nu} / \dot{\nu}$ $\thicksim$ 7.1$\times 10^{-4}$. It decreased back towards initial value after glitch. There was an increase in $\ddot{\nu}$ (see Table~\ref{Tab:table3}), which was caused by the post$\textrm{-}$glitch recovery.

\section{Discussion}
\label{sect:discussion}

As of April 2018, PSR B1737$-$30 has been observed to exhibit 36 glitches, including this one. Information of glitch epoch, glitch interval, $\Delta \nu/ \nu$ and reference for all the 36 glitches are listed in Table~\ref{Tab:table4}. The fractional increase of $\Delta \nu/ \nu$ widely ranges from 7$\times 10^{-10}$ to $\thicksim$ 2.66$\times 10^{-6}$. The $\Delta \nu/ \nu$ of the glitch detected by the TMRT is about 8.39$\times 10^{-7}$, making it to be the 4th largest glitch known in this pulsar.

\begin{table}[ht]
\footnotesize{
\caption{Information of 36 glitches in PSR B1737$-$30.}
\label{Tab:table4}
\centering
\begin{tabular}{c l l l c}
\hline
\hline
\multicolumn{1}{c}{Number} &
\multicolumn{1}{c}{Epoch ($T_{\rm i}$)} &
\multicolumn{1}{c}{Interval ($\Delta T_{\rm i}$)} &
\multicolumn{1}{c}{$\Delta \nu / \nu $} &
\multicolumn{1}{c}{Reference}
\\
 &
\multicolumn{1}{c}{(MJD)}  &
\multicolumn{1}{c}{(d)} &
\multicolumn{1}{c}{(10$^{-9}$)} &
\multicolumn{1}{c}{ }\\
\hline
1   & 46991(19)	    & 290(19)     &	421(4)        & 7 \\
2   & 47281(2)  	    & 51(16)      &	33(5)         & 1 \\
3   & 47332(16)	    & 126(16)     &	7(5)          & 1 \\
4   & 47458(2)      & 212.2(20)   &	30(8)         & 1 \\
5   & 47670.2(2)    & 487.8(10)   &	600.9(6)      & 1 \\
6   & 48158(1) 	    & 33.7(10)    &	10(1)         & 7 \\
7   & 48191.69(0)   & 26.3(20)    &	659(7)        & 7 \\
8   & 48218(2) 	    & 213.3(20)   &	48(10)        & 2 \\
9   & 48431.3(4)    & 616.2(6)    &	16(2)         & 7 \\
10  & 49047.5(5)  	& 191.6(5)    &	17(1)         & 7 \\
11  & 49239.07(2)	& 212.6(4)    &	169.7(2)      & 7 \\
12  & 49451.7(4)	    & 92.2(4)     &	9.5(5)        & 4 \\
13  & 49543.93(8)	& 1030.62(8)  &	3.0(6)        & 4 \\
14  & 50574.5497(4)	& 367.0685(4) &	439.3(2)      & 4 \\
15  & 50941.6182(2)	& 743(21)     &	1443.0(3)     & 3,4 \\
16  & 51685(21)	    & 142(21)     &	0.7(4)        & 5 \\
17  & 51827(2)    	& 221(9)      &	0.9(3)        & 7 \\
18  & 52048(9)   	& 197(9)      &	2(3)          & 7 \\
19  & 52245(2)   	& 21(2)       &	4(1)          & 7 \\
20  & 52266.0(2) 	& 81.7(2)     &	16(1)         & 7 \\
21  & 52347.66(6)	& 228.3(30)   &	152(2)        & 6 \\
22  & 52576(3)   	& 203.7(30)   &	0.9(2)        & 7 \\
23  & 52779.70(4)	& 79.08(5)    &	1.7(2)        & 7 \\
24  & 52858.78(3)	& 83.7(1)     &	18.6(3)       & 7 \\
25  & 52942.5(1) 	& 81.0(1)     &	20.2(2)       & 7 \\
26  & 53023.52(0)	& 450.04(1)   &	1850.9(3)     & 6,7 \\
27  & 53473.56(1)	& 976.63(1)   &	0.8(2)        & 7 \\
28  & 54450.19(1)	& 245(0)      &	45.9(3)       & 7 \\
29  & 54695.19(2)	& 115.7(1)    &	3.0(2)        & 7 \\
30  & 54810.9(1) 	& 117.7(1)    &	5.2(3)        & 7 \\
31  & 54928.6(1) 	& 291(14)     &	2.3(2)        & 7 \\
32  & 55220(14)	    & 2076(14)    &	2664.50(15)   & 8 \\
33  & 57296.5(9)	    & 49.5(11)    &	1.30(4)       & 9 \\
34  & 57346.0(6)	    & 153.4(6)    &	1.94(2)       & 10 \\
35  & 57499.371(4)	& 732.0(4)    &	227.29(3)     & 10 \\
36  & 58232.4(4)  	&             &	838.7(5)     & this work\\

\hline
\end{tabular}
\\
Note: These parameters are referenced from ATNF pulsar database. The glitch interval $\Delta T_{\rm i}$ is defined as: $\Delta T_{\rm i}$ = $T_{\rm i+1}$ $-$ $T_{\rm i}$.
Reference: 1. \cite{ml90}; 2. \cite{sl96}; 3. \cite{ura02}; 4. \cite{klgj03}; 5. \cite{js06}; 6. \cite{zwm+08}; 7. \cite{elsk11}; 8. \cite{ymh+13}; 9. \cite{jbb+15}; 10. \cite{jbb+16}
}
\end{table}

The parameter $A_{\rm g}$ is defined as the mean fractional frequency variation per year caused by glitches
\begin{equation}
  A_{\rm g} =\frac{ 1 }{ T_{\rm g}}\sum{\frac{\Delta \nu}{\nu}},
\label{eq:equation6}
\end{equation}
where the $\sum{(\Delta \nu /\nu)}$ is the sum of fractional increments in $\nu$ of all glitches during the interval $T_{\rm g}$ \citep{ml90}. The $A_{\rm g}$ depends on not only how frequently the glitches occurred, but also the size of glitches. This makes it to be a good indicator of glitch events on long time$\textrm{-}$scale, as it is mainly dominated by large glitch and insensitive to small glitch which is sometimes difficult to be distinguished from timing noise. The glitch activity parameter $A_{\rm g}$ of PSR B1737$-$30 is about 3.17$\times 10^{-7}$ yr$^{-1}$.

\begin{table}[ht]
\footnotesize{
\caption{Glitch activity parameters of eight frequently glitching pulsars.}
\label{Tab:table5}
\centering
\begin{tabular}{l c l l l l c}
\hline
\hline
\multicolumn{1}{c}{Name} &
\multicolumn{1}{c}{$N_{\rm g}$} &
\multicolumn{1}{c}{$\sum{(\Delta \nu/ \nu)}$} &
\multicolumn{1}{c}{Time span} &
\multicolumn{1}{c}{$T_{\rm g}$} &
\multicolumn{1}{c}{$A_{\rm g}$} &
\multicolumn{1}{c}{$\tau_{\rm c}$}
\\
 &
 &
\multicolumn{1}{c}{(10$^{-9}$) } &
\multicolumn{1}{c}{(MJD)} &
\multicolumn{1}{c}{(yr)}  &
\multicolumn{1}{c}{(10$^{-9}$ yr$^{-1}$)} &
\multicolumn{1}{c}{(kyr)}\\
\hline
 B0531$+$21    & 25 & 977.7(8)   & 40491.8(3)$\textrm{-}$58064.555(3)   & 48.14(0) & 20.31(2)  & 1.26 \\
 J0537$-$6910  & 23 & 6614(23)   & 51285.7(8.6)$\textrm{-}$53951.2(1.5) & 7.26(0)  & 911(3)    & 4.93 \\
 B0833$-$45    & 19 & 34811(20)  & 40280(4)$\textrm{-}$56922(3)         & 45.59(1) & 743.4(4)  & 11.3 \\
 B1338$-$62    & 23 & 11226(6)   & 47989(21)$\textrm{-}$55088(16)       & 19.45(7) & 577.2(3)  & 12.1 \\
 B1737$-$30    & 36 & 9765(3)    & 46991(19)$\textrm{-}$58232.4(4)      & 30.80(5) & 317.1(1)  & 20.6 \\
 J0631$+$1036  & 14 & 5082.9(6)  & 50183.5(2)$\textrm{-}$55702(3)       & 15.12(1) & 336.17(4) & 43.6 \\
 B1758$-$23    & 10 & 2118(1)    & 46907(21)$\textrm{-}$55356(3)        & 23.15(6) & 91.68(6)  & 58.3 \\
 B1822$-$09    & 12 & 242.7(8)   & 49615(8)$\textrm{-}$54115.78(4)      & 12.33(2) & 19.68(6)  & 232  \\
\hline
\end{tabular}
\\
Note: The $N_{\rm g}$ is the number of glitches. The $\sum{(\Delta \nu/ \nu)}$ is the cumulative fractional glitch size of each pulsar, its error is taken as the variance of errors from each $\Delta \nu / \nu$ for every pulsar. The $T_{\rm g}$ equals the interval of time span in years.
}
\end{table}

Beside PSR B1737$-$30, there are other pulsars with various glitches reported. Parameters of eight pulsars with at least 10 glitches record are listed in Table~\ref{Tab:table5}. The glitch size $\Delta \nu/ \nu$, time span and $\tau_{\rm c}$ are referenced from ATNF pulsar database. The parameter $T_{\rm g}$ is calculated as the length of time span in years. This table is listed in sequence of $\tau_{\rm c}$. From last two columns, the $A_{\rm g}$ generally decreases when the $\tau_{\rm c}$ increases for pulsars with $\tau_{\rm c}$ $\geq$ 4 kyr. But the Crab pulsar (B0531$+$21) \citep{sr68} is an exception. Most of the 25 glitches detected in this pulsar are small or middle$\textrm{-}$size glitches. Only two relatively large glitches are measured to be $\Delta \nu/ \nu$ about 2.14$\times 10^{-7}$ and 4.8$\times 10^{-7}$ on MJD $\thicksim$ 53067.1 and 58064.6, respectively. There is a possible explanation for relative weak glitch events of the Crab. Very young pulsars with $\tau_{\rm c}$ smaller than 2 kyr have higher temperature which reduces the effect of pinning force and makes it more easier for superfluid vortices to move outward. So that, angular momentum are transferred more smoothly from superfluid to the crust, causing the glitch size more likely to be small \citep{ml90}.

\begin{figure}[h]
\begin{center}
\begin{tabular}{cc}
\resizebox{0.75\hsize}{!}{\includegraphics[angle=0]{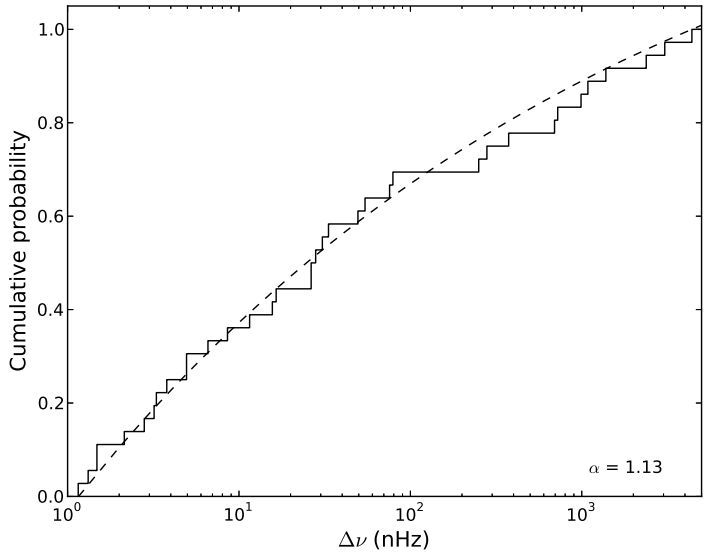}}\\
\end{tabular}
\end{center}
\caption{ Cumulative frequency increment distribution of PSR B1737$-$30, and the power law fit given by Equation~\ref{eq:equation7} with index $\alpha$ = 1.13 (dashed curve).}
\label{fg:msRAA-2018-0260R1fig5}
\end{figure}

If glitches are resulted from the avalanche process, their sizes follow a power law distribution in the form of:

\begin{equation}
  P(\Delta \nu) =\frac{ \Delta \nu^{(1- \alpha)} - \Delta \nu_{\rm min}^{(1- \alpha)} }{ \Delta \nu_{\rm max}^{(1- \alpha)} - \Delta \nu_{\rm min}^{(1- \alpha)}}\qquad,
\label{eq:equation7}
\end{equation}
where $\alpha$, $\Delta \nu_{\rm max}$ and $\Delta \nu_{\rm min}$ are the power law index, the maximum and minimum frequency jumps, respectively \citep{mpw08}. The Kolomogorov$\textrm{-}$Smirnov (K$\textrm{-}$S) test statistic $D$ and its associated $P_{\rm K \textrm{-}S}$ are referenced to measure the agreement between data and power law fit. The parameter $D$ is the maximum difference between two data sets, and the $P_{\rm K \textrm{-}S}$ means the probability that two sets of data follow the same distribution, equally implies how well the glitch size distribution is described by the power law. For PSR B1737$-$30, the cumulative glitch size distribution based on glitches before MJD 53190 (July 4, 2004) was fitted by a power law function \citep{mpw08}. According to their calculation, the $P_{\rm K \textrm{-}S}$ is 0.992 relative to the best fitted $\alpha$ = 1.1, suggesting a good description for cumulative glitch size distribution by the power law. As ten more glitches occurred in PSR B1737$-$30 after MJD 53190, it is necessary to fit the cumulative glitch size distribution again to test whether it still follows the power law distribution. The power law fit was performed on 36 glitches listed in Table~\ref{Tab:table4}. The final $\alpha$ value 1.13$\pm$0.03 was chosen when the $P_{\rm K \textrm{-}S}$ became maximum. The relative error was estimated as the range corresponding to the $P_{\rm K \textrm{-}S}$ $\geq$ 0.985 confidence. The K$\textrm{-}$S statistic was calculated to be $D$ = 0.07 with an associated $P_{\rm K \textrm{-}S}$ = 0.9996. This implies that the glitch size distribution of PSR B1737$-$30 still well obeys the power law distribution, although more glitches with different sizes occurred. Cumulative glitch size distribution of PSR B1737$-$30 is shown in Fig~\ref{fg:msRAA-2018-0260R1fig5}, together with the power law fit described in Equation~\ref{eq:equation7} (dashed curve).

\begin{figure}[h]
\begin{center}
\begin{tabular}{cc}
\resizebox{0.75\hsize}{!}{\includegraphics[angle=0]{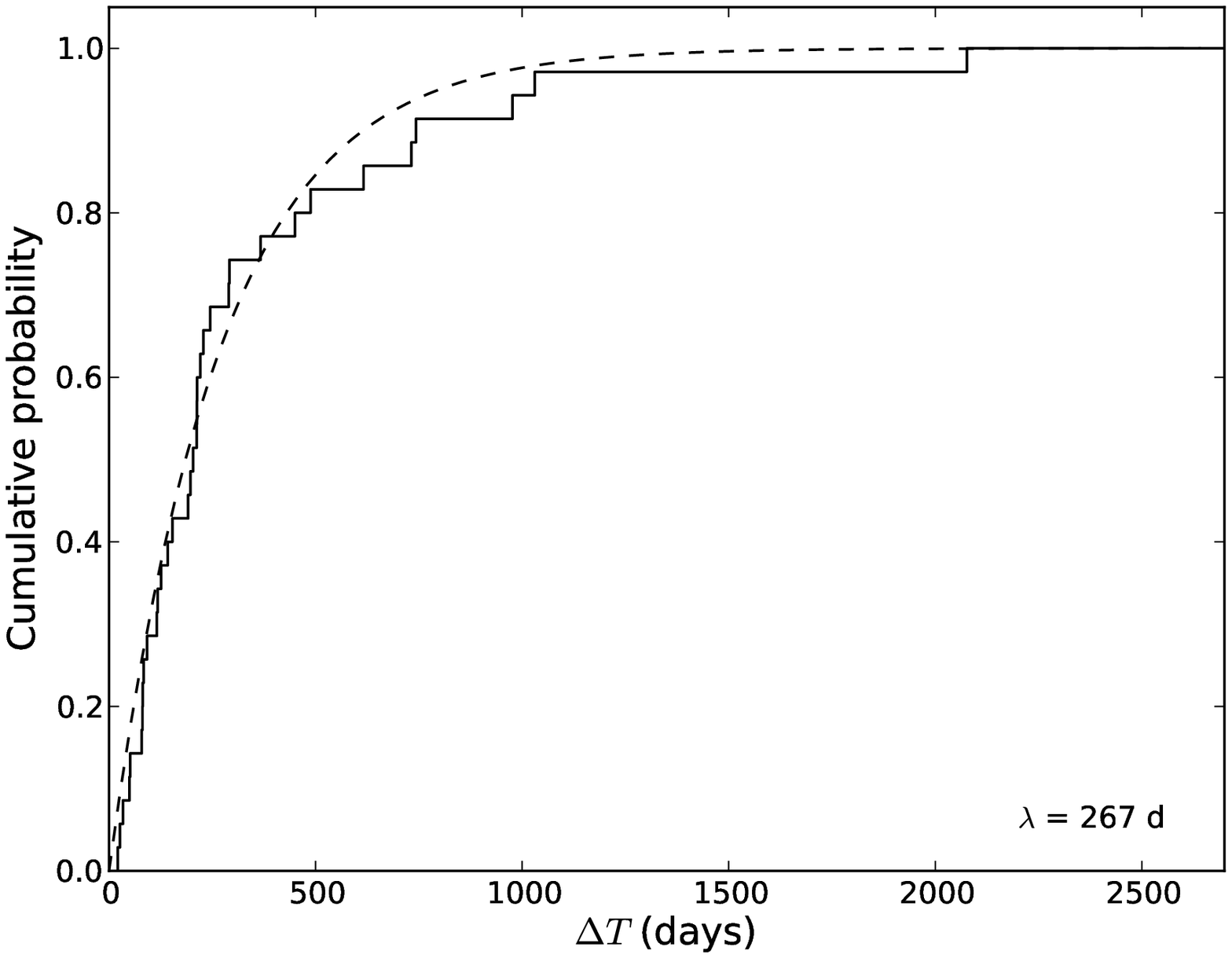}}\\
\end{tabular}
\end{center}
\caption{Cumulative $\Delta T$ distribution of PSR B1737$-$30 corresponding to 35 waiting times and the Poisson model fit in Equation~\ref{eq:equation8} with $\lambda$ = 267 d (dashed curve). Errors of $\Delta T$ are ignored.}
\label{fg:msRAA-2018-0260R1fig6}
\end{figure}

Pulsar glitches are statistically independent if they are caused by an avalanche process. This can be explained by a system in the state of Self Organised Criticality (SOC). The system is described as a combination of many metastable reservoirs separated from each other by relaxed regions. Stress cumulated in every reservoir is released out during one avalanche process. The following avalanche happens randomly and is typically far from previous one. No interference is found between two adjacent avalanches \citep{jen98}. The interval between two adjacent glitches is defined as waiting time $\Delta T$. Based on the statistical independence of glitches, considering that the system is driven by the local nearest force at a mean rate, the avalanche model predicts that $\Delta T$ follows the Poisson statistics. So the distribution of $\Delta T$ can be described by a Poissonian probability density function as:
\begin{equation}
  p(\lambda,t) =\lambda^{-1}e^{-t/\lambda},
\label{eq:equation8}
\end{equation}
where $\lambda$ is the mean waiting time. \cite{mpw08} fitted waiting time distribution of PSR B1737$-$30 over glitches before MJD 53190 with the best fitted $\lambda$ $\thicksim$ 242 d. They proposed that the $\lambda$ is not expected to vary obviously in a range of forty years. We calculated waiting times based on all detected glitches, got the best fitted $\lambda$ = 267 d of the Poisson model. This means a good consistence with the result calculated in \cite{mpw08}. The cumulative waiting time distribution and the best model fit (dashed curve) are plotted in Fig~\ref{fg:msRAA-2018-0260R1fig6}. The K$\textrm{-}$S statistic between data and the Poisson model was calculated to be $D$ = 0.086 with the associated $P_{\rm K \textrm{-}S}$ = 0.999, implying a good agreement between data and model.

\begin{figure}[h]
\begin{center}
\begin{tabular}{cc}
\resizebox{0.5\hsize}{!}{\includegraphics[angle=0]{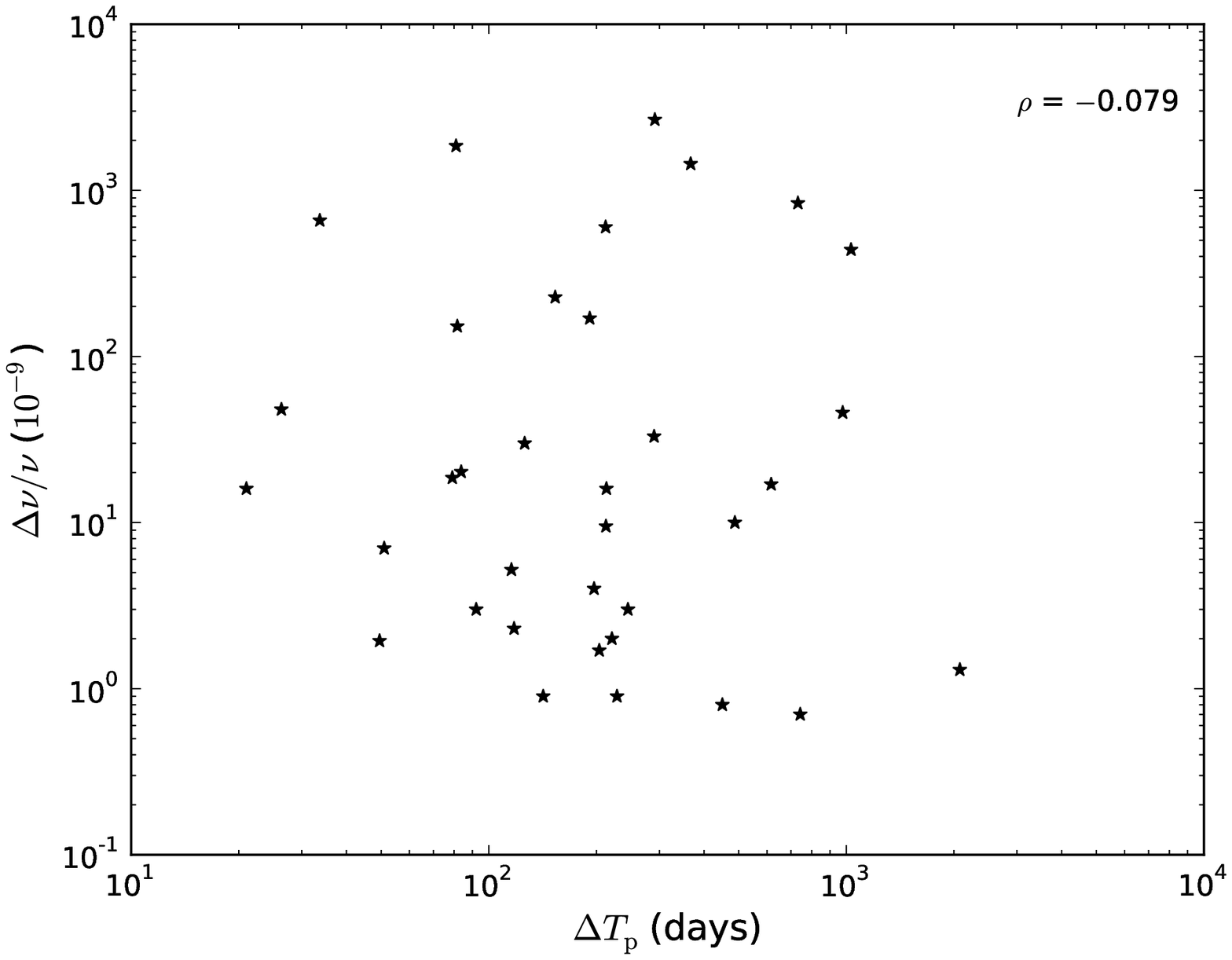}}&
\resizebox{0.5\hsize}{!}{\includegraphics[angle=0]{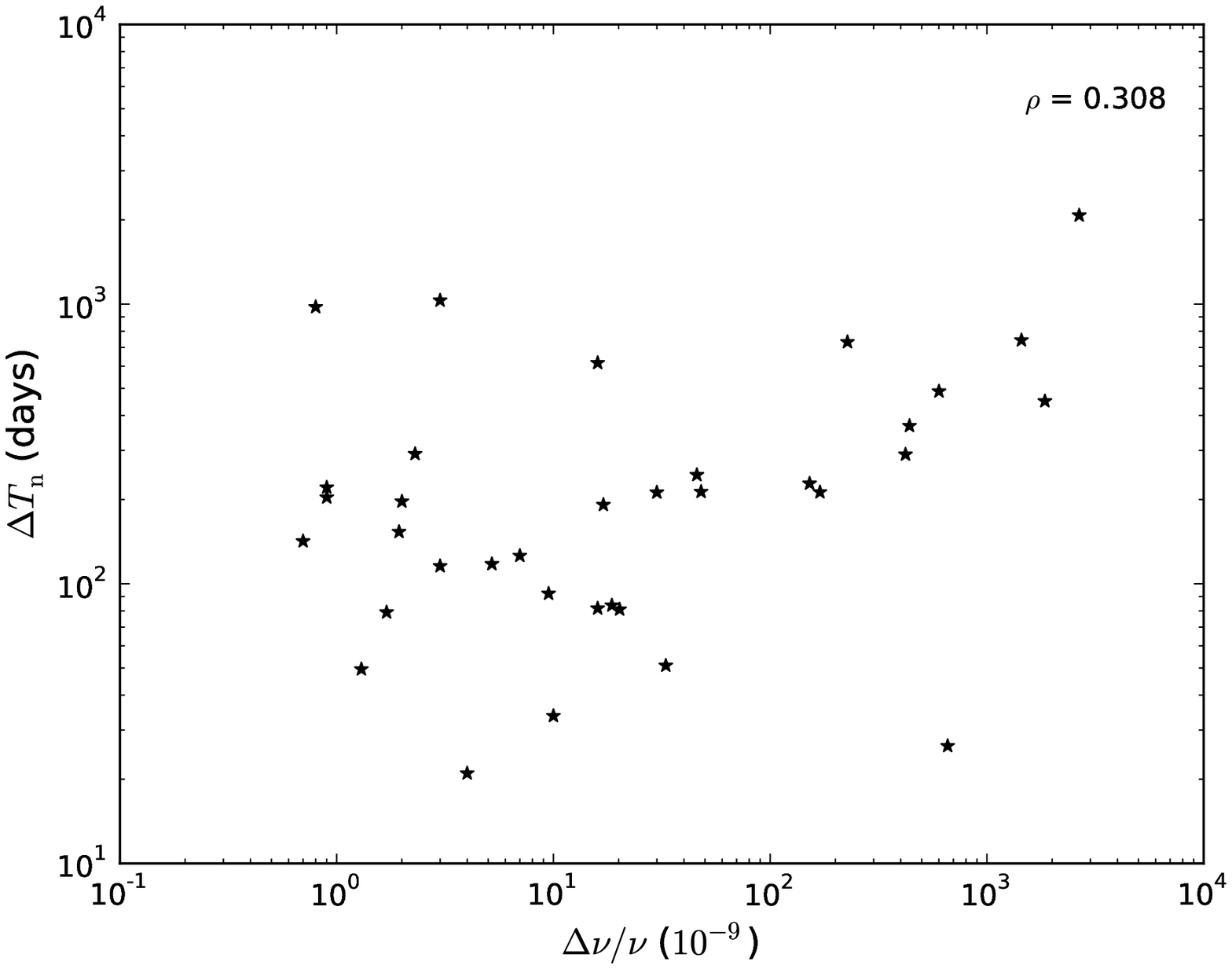}}\\
\end{tabular}
\end{center}
\caption{ The relation between $\Delta T_{\rm p}$ and glitch size $\Delta \nu/ \nu$ (left panel) together with the relation between $\Delta \nu/ \nu$ and $\Delta T_{\rm n}$ (right panel). Errors of $\Delta \nu/ \nu$ and $\Delta T$ are ignorable for most points. }
\label{fg:msRAA-2018-0260R1fig7}
\end{figure}

There are 36 glitches record on PSR B1737$-$30, making it a good sample to test the correlation between $\Delta T$ and glitch size ($\Delta \nu/ \nu$). In the left panel of Fig~\ref{fg:msRAA-2018-0260R1fig7}, the glitch size is plotted against the $\Delta T$ since previous glitch ($\Delta T_{\rm p}$). As data points are so obviously diffuse with the Spearman Rank correlation coefficient $\rho$ = $-$0.079, no correlation is found between them. However, a weak correlation ($\rho$ = 0.4) between the $\Delta T_{\rm p}$ and the glitch sizes of the Crab was proposed that large glitches are more possible to take place after long glitch intervals \citep{slsw+18}. It is necessary to mention that the large glitch at MJD 58064 dominates this correlation of the Crab, and few small glitches of the Crab happened after long waiting times. This correlation could possibly result from the so called ``reservoir effect". In this scenario, the angular momentum are firstly stored then completely released into the crust during a glitch \citep{hm15,slsw+18}. The $\Delta T$ before next glitch ($\Delta T_{\rm n}$) is plotted against glitch size in the right panel of Fig~\ref{fg:msRAA-2018-0260R1fig7}. There is little correlation ($\rho$ = 0.308) between this two terms that the $\Delta T_{\rm n}$ prefers to be long after a large glitch. An apparent correlation ($\rho$ = 0.931) was demonstrated between glitch size and $\Delta T_{\rm n}$ in PSR J0537$-$6910 too \citep{aeka18,mhf18}. It is much stronger than that in PSR B1737$-$30, but is not universal.

\section{Conclusion}
\label{sect:conclusion}

We present one large glitch in PSR B1737$-$30 detected with the TMRT around MJD 58232.4. PSR B1737$-$30 is the most frequently glitching pulsar with 36 glitches already detected. The glitch at MJD 58232.4 underwent a frequency increment of $\thicksim$ 1.38$\times 10^{-6}$ Hz, corresponding to the fractional increase of $\Delta \nu/ \nu$ $\thicksim$ 8.39$\times 10^{-7}$. The parameter $A_{\rm g}$ is a good indicator of glitch events. For PSR B1737$-$30, the value of $A_{\rm g}$ is about 3.17$\times 10^{-7}$ yr$^{-1}$. Based on the statistics of glitches in eight pulsars with at least 10 glitch events, we find a correlation between $A_{\rm g}$ and characteristic age $\tau_{\rm c}$. For pulsars whose $\tau_{\rm c}$ are greater than 4 kyr, the $A_{\rm g}$ generally decreases when the $\tau_{\rm c}$ becomes larger. Glitch size distribution of PSR B1737$-$30 follows the power law distribution with index of 1.13. The distribution of $\Delta T$ obeys the Poissonian probability density function with best fitted $\lambda$ = 267 d. No correlation is found between glitch size and the waiting time $\Delta T_{\rm p}$, but the $\Delta T_{\rm n}$ after large glitch is more likely to be long. Since pulsar glitches differ a lot from one to another even in the same pulsar, a larger glitch sample is valuable for characterizing glitch activities.

\section*{Acknowledgements}

We would like to express our appreciation to professor R.N.Manchester and anonymous reviewers for their good suggestions on this work. This work was supported in part by the National Natural Science Foundation of China (Grants NO. U1631122, NO. 11403073 and NO. 11633007), the strategic Priority Research Program of Chinese Academy of Sciences (Grant NO. XDB23010200), the Knowledge Innovation Program of the Chinese Academy of Sciences (Grant NO. KJCX1-YW-18) and the National Key R$\&$D Program of China (Grant NO. 2018YFA0404602). The hard work of all member of the TMRT team is vital for high quality observation data used in this paper.

\label{lastpage}

\clearpage

\end{document}